\documentclass[reprint,twocolumn,amsmath,amssymb,aps,prl,]{revtex4-2}

\usepackage{graphicx, float}
\usepackage{dcolumn}
\usepackage{bm, braket, commath}

\newcommand{\BK}{Baym–Kadanoff}
\newcommand{\LW}{Luttinger–Ward}
\newcommand{\MT}{Matsubara}
\newcommand{\MH}{Mattheiss-Hamann}

\newcommand{\Tc}{T_{\text{c}}}
\newcommand{\chiCDW}{\chi^{\text{CDW}}}
\newcommand{\chiCDWmax}{\chiCDW_{\max}}
\newcommand{\irrchi}{\chi_{0}}

\newcommand{\imgI}{\mathrm{i}}
\newcommand{\kvec}{\mathbf{k}}
\newcommand{\qvec}{\mathbf{q}}
\newcommand{\qmax}{\qvec_{\max}}
\newcommand{\dcdw}{\delta_{\text{c}}}

\begin{document}

\title{Commensurate to Incommensurate Transition of Three Dimensional Charge Density Waves}

\author{Hao Wang}
\affiliation{School of Physics, Peking University, Beijing 100871, People’s Republic of China}

\author{Qiang Luo}
\email{luoqiang@pku.edu.cn}
\affiliation{School of Physics, Peking University, Beijing 100871, People’s Republic of China}

\author{Ji Chen}
\email{ji.chen@pku.edu.cn}
\affiliation{School of Physics, Peking University, Beijing 100871, People’s Republic of China}
\affiliation{Interdisciplinary Institute of Light-Element Quantum Materials and Research Center for Light-Element Advanced Materials,Peking University, Beijing 100871, P. R. China
}
\affiliation{Frontiers
Science Center for Nano-Optoelectronics, Peking University, Beijing 100871, P. R. China}

\date{September 2022}

\begin{abstract}
Charge density wave (CDW) is a widely concerned emergent phenomenon in condensed matter physics.
To establish a systematic understanding of CDW, we develop a diagrammatic self-consistent-field approach for cubic Holstein model employing fluctuation exchange approximation, and explore the emergence and transition of three-dimensional CDWs.
Commensurate CDW (c-CDW) locked at $(\pi,\pi,\pi)$ is favored near half-filling, and the transition temperature is predicted around half of the nearest-neighbor hopping.
Large hole doping leads to a suppression of CDW transition temperature and the emergence of incommensurate CDW (i-CDW), which is evidenced by a drifting of the ordering vector away from $(\pi,\pi,\pi)$ towards $(\pi,\pi,0)$. 
Phonon frequency significantly impacts the transition temperature and the phase boundary between c-CDW and i-CDW, and the optimal frequency for enlarging the CDW regime is also predicted near half of the nearest-neighbor hopping.
These new theoretical results provide a systematic understanding of CDW and a fresh perspective on emergent phenomena dominated by electron-phonon interaction.
\end{abstract}

\maketitle


Charge density wave (CDW), a collective phenomenon that modulates charge density and lattice, is ubiquitous in different dimensions and plays a pivotal role in condensed matter physics since Peierls' proposal in 1955 \cite{Peierls2001, hwang2024charge}.
CDWs of different dimensions and orderings have been continuously revealed in experiments and predicted theoretically during the past decades \cite{neto2001charge, tang2019three, qin2020theory, liang2021three, liu2024charge, nhat2024three}.
The new phases of CDWs have greatly deepened our understanding of emergent phenomena of electrons in condensed matter.
A notable example is the relation between CDW and superconductivity, around which there are a lot of debates about whether they coexist or compete with each other \cite{Rice1979, sykora2009coexistence, Costa2018}.
Recently the transition from commensurate to incommensurate CDW bears an unusual M-shaped double superconducting dome in kagome material \cite{yu2021unusual, chen2021double, feng2023commensurate}.

Theoretically, to understand the emergence of CDW, an often employed theoretical model is the Holstein model that combines single-band tight-binding model with explicit electron-phonon (el-ph) interaction, described by the following Hamiltonian
\begin{equation}
\begin{split}
    H = & -t\sum_{\braket{i,j}\sigma} ( c^{\dagger}_{i\sigma}c_{j\sigma} + h.c. ) - \mu\sum_{i\sigma} n_{i\sigma} \\
    & + g\sum_{i\sigma} ( b^{\dagger}_i + b_i ) n_{i\sigma} + \Omega\sum_i b^{\dagger}_i b_i
\end{split}
\end{equation}
where $\braket{i,j}$ traverses the nearest-neighbor pairs, $c_{i\sigma}^\dagger$ creates an electron of spin $\sigma$ on site $i$, $b_i^\dagger$ creates a bare phonon of frequency $\Omega$ on site $i$, $t$ is the nearest-neighbor hopping, $\mu$ is the chemical potential used to tuning band filling $n$ or say hole doping $\delta$, and $g$ is the el-ph coupling constant.
Despite its simple form, exact analytical solution of the Holstein model does not exist, except for some extreme cases such as the two-site problem \cite{Berciu2007, Han2022, Zhang2009, marsiglio2022impact} and the atomic limit \cite{mahan1981many}. 
Previous numeric studies on Holstein model mostly focus on 1D \cite{Hirsch1983, marsiglio1995pairing, wellein1998self, jeckelmann1998density, Jeckelmann1999, Cataudella2000, hohenadler2004quantum, cheng2008unified, Greitmann2015, mitric2022spectral, Jansen2023}, 2D \cite{scalettar1989competition, marsiglio1990pairing, noack1991charge, noack1993green, zheng1997charge, jeckelmann1998density, sykora2009coexistence, weber2018two, esterlis2018breakdown, dee2019temperature, costa2020phase, han2020strong, paleari2021quantum, bradley2021superconductivity, karakuzu2022stripe}, or infinite dimension \cite{freericks1993holstein, freericks1995competition, benedetti1998holstein, meyer2002gap, park2019dynamical, perepelitsky2019band}.
Solving 3D Holstein model is a much more challenging problem.
So far very few studies have been reported, with e.g.\ an exact-diagonalization based Monte Carlo (ED-MC) approach with travelling cluster approximation (TCA) \cite{Kumar2005} and a Langevin-dynamics based determinant quantum Monte Carlo (DQMC) method \cite{Cohen2020}.
Due to computational complexity, these studies are limited to calculations either within classical phonon approximation \cite{Kumar2005} or at half-filling \cite{Cohen2020}.

Here we report a study of 3D CDW in cubic Holstein model, based on the development of an efficient diagrammatic self-consistent field (SCF) method.
The overall computational framework and implementation details are presented in the supporting information (SI Sec.~1).
In the following, the nearest-neighbor hopping $t$ is taken as the energy unit.
The el-ph coupling constant $g=1.10668191970032$ is set according to Ref.~\onlinecite{Cohen2020}.
We use the lattice of $L=12$ and truncate the \MT{} frequency to $80$.
We tune doping level $\delta$ in between $0\%$ and $40\%$ where the truncation of the \MT{} frequency doesn't cause evident bias.

The key part of this computational framework is fluctuation exchange (FLEX) approximation \cite{Bickers1989, Pao1999}, a conserving approximation generated via \BK{} formalism \cite{baym1961conservation, baym1962self}.
In \BK{} formalism, a functional about Green function $\Phi[G]$, also called \LW{} functional \cite{luttinger1960ground}, is constructed to preserve conservation laws for particle number, energy, momentum and angular momentum, and self-energy is a functional derivative expressed as $\Sigma[G]=\delta\Phi[G]/\delta G$.
After $\Phi[G]$ diagrammatically expanded, one can retain part of diagrams to approximate it. 
In FLEX approximation, the diagrams symmetrically describing exchange of fluctuations in particle-hole and particle-particle channels are retained.
After FLEX-based SCF is converged, one obtains single-particle self-energy $\Sigma(k)$ and Green function $G(k)$ at finite temperature $T$, where $k=(\kvec{}, \imgI{}\omega_n)$ and $q=(\qvec{}, \imgI{}\nu_m)$ for fermion and boson respectively, and \MT{} frequencies at temperature $T$ are denoted as $\omega_n=(2n+1)\pi T$ and $\nu_m=2m\pi T$ respectively with integers $n$ and $m$.


The CDW phase transition temperature $\Tc$ is determined by the divergence of CDW susceptibility
\begin{equation}
    \chiCDW(\qvec{}) = \frac{1}{N} \int_0^{\frac{1}{T}} \dif\tau 
    \braket{\hat{\rho}_{\qvec{}}(\tau) \hat{\rho}_{\qvec{}}^{\dagger}(0)}_c
\end{equation}
where
\begin{equation}
    \hat{\rho}_{\qvec{}}(\tau) = 
    \sum_{i\sigma} e^{-\imgI{}\qvec{}\cdot\mathbf{R}_i} 
    c^{\dagger}_{i\sigma}(\tau)c_{i\sigma}(\tau)
\end{equation}
and the connected charge density correlation is defined as $\braket{\hat{\rho} \hat{\rho}^{\dagger}}_c = \braket{\hat{\rho} \hat{\rho}^{\dagger}} - \braket{\hat{\rho}}\braket{\hat{\rho}^{\dagger}}$.
As the charge density correlation becomes more and more long-range-like, the CDW susceptibility peaks at an ordering vector $\qmax$.
Once the long-range charge density correlation is established, i.e.\ CDW emerges, the peak of $\chiCDW$ at $\qmax$ diverges.
Within Migdal approximation \cite{dee2019temperature}, the CDW susceptibility is simplified to
\begin{equation} \label{eq:chiCDWMigdal}
    \chiCDW(\qvec{}) = \frac {\irrchi(\qvec{}, 0)} {1 - U_p \irrchi(\qvec{},0)}
\end{equation}
where $\irrchi$ is the irreducible charge susceptibility defined as
\begin{equation}
    \irrchi(q) = -\frac{2T}{N} \sum_k G(k)G(k+q)
\end{equation}
and $U_p = 2g^2/\Omega$ is the phonon-mediated on-site effective attraction between electrons.
For convenience, we'll denote $U_p \irrchi(\qvec{},0)$ as $\lambda(\qvec{})$.

According to Eq.~(\ref{eq:chiCDWMigdal}), CDW phase transition happens when the denominator approaches zero, i.e.\ $\lambda(\qvec{})$ approaches unity at some ordering vector $\qmax$.
So the determination of $\Tc$ is equivalent to solving
\begin{equation} \label{eq:determine_Tcdw}
    \max_{\qvec{}} \set{\lambda(\qvec{}; \Tc)} = 1
\end{equation}
When lowering temperature, charge density correlation is enhanced and $\lambda_{\max}$ becomes larger, even surpassing unity.
So we can use the bisection method to work out $\Tc$.
To determine whether a CDW phase is commensurate (c-CDW) or incommensurate (i-CDW), the ordering vector $\qmax$ is computed.
We track the distribution of $\chiCDW$ in reciprocal space when lowering temperature until $\lambda_{\max}$ surpasses unity to determine $\qmax$.
If $\qmax$ is locked at $(\pi, \pi, \pi)$ when lowering temperature, then the CDW is commensurate.
Otherwise, if $\qmax$ drifts away from $(\pi, \pi, \pi)$, the CDW is regarded as incommensurate with $\qmax=(\pi, \pi, \kappa\pi)$ where $0<\kappa<1$.
It should be noted that the coarse momentum space precludes the precise determination of the ordering vector for i-CDW, but it is evident to suggest a transition to i-CDW.



\begin{figure}[t]
    \centering
    \includegraphics[width=0.48\textwidth]{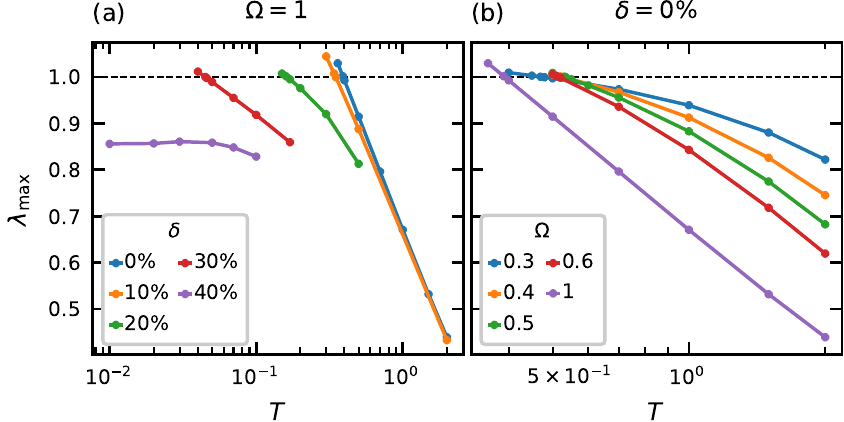}
    \caption{(a) and (b) shows the dependency of $\lambda_{\max}$ on temperature at $\Omega=1$ with $\delta=0,10,20,30,40$ percent and half-filling with $\Omega=0.3,0.5,1$ respectively.
    When lowering temperature, $\lambda_{\max}$ is enlarged and can even cross over unity, where $\Tc$ is determined.}
    \label{fig:determine_Tcdw}
\end{figure}

Figure~\ref{fig:determine_Tcdw} shows the dependency of $\lambda_{\max}$ on $T$ and the determination of $\Tc$.
When $\Omega=1$ with $\delta = 0, 10, 20, 30, 40$ percent [Fig.~\ref{fig:determine_Tcdw}(a)], $\lambda_{\max}$ and hence $\chiCDWmax$ are nearly independent of $\delta$ at the temperature much higher than $\Tc$, but are monotonically suppressed when enlarging $\delta$ at lower temperature.
According to Eq.~(\ref{eq:determine_Tcdw}), $\Tc$ is determined as $0.39$, $0.35$, $0.16$ and $0.045$ for $\delta=0,10,20,30$ percent respectively.
For $\delta=40\%$, we do not observe any CDW phase transition down to $T=0.01$.
At half-filling with $\Omega = 0.3, 0.4, 0.5, 0.6, 1$ [Figure~\ref{fig:determine_Tcdw}(b)], $\lambda_{\max}$ and hence $\chiCDWmax$ are monotonically suppressed when enlarging $\Omega$ at the temperature higher than $\Tc$.
According to Eq.~(\ref{eq:determine_Tcdw}), $\Tc$ is determined as $0.48$, $0.53$, $0.533$, $0.52$ and $0.39$ for $\Omega = 0.3, 0.4, 0.5, 0.6, 1$ respectively at half-filling.


\begin{figure*}[t]
    \centering
    \includegraphics[width=\linewidth]{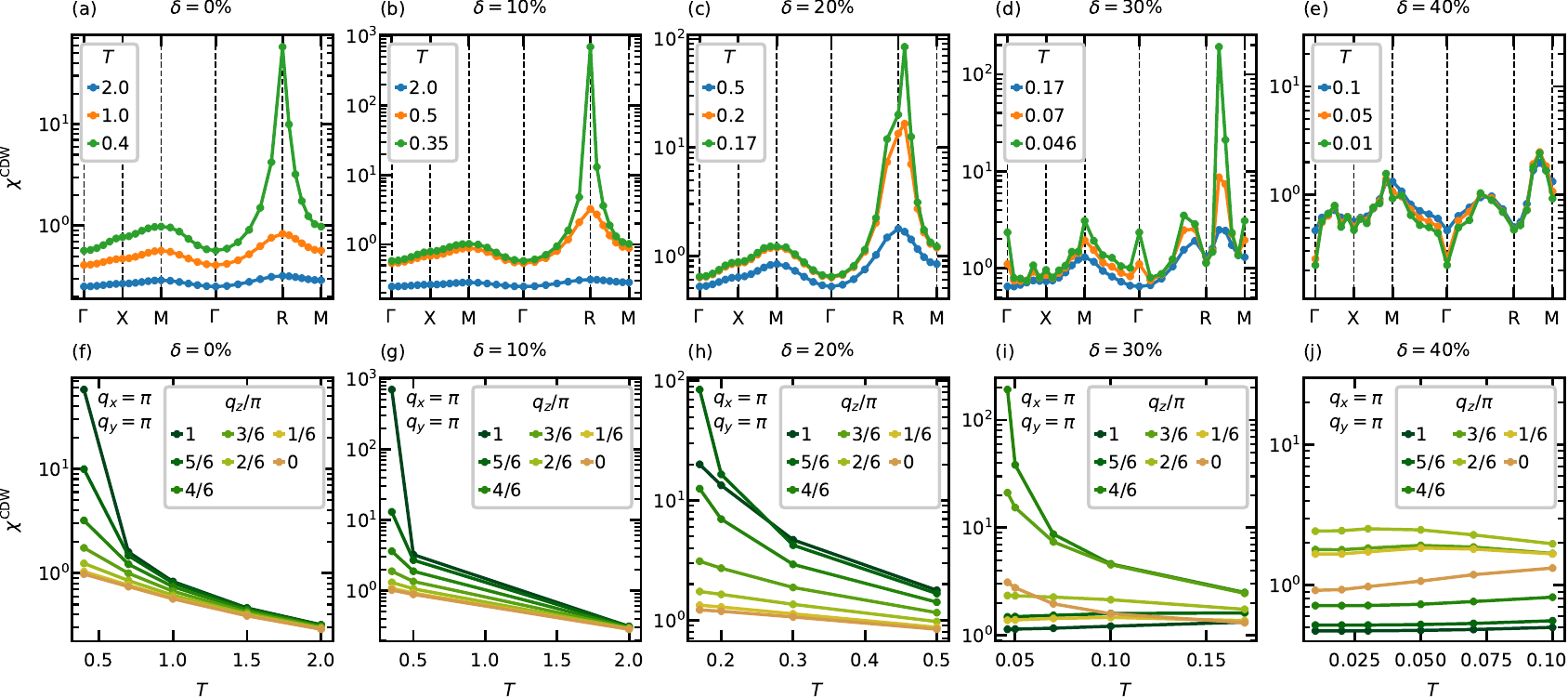}
    \caption{$\chiCDW$ above $\Tc$ at $\Omega=1$ with $\delta = 0, 10, 20, 30, 40$ percent.
    (a-e) show the distribution of $\chiCDW$ on path $\Gamma$-X-M-$\Gamma$-R-M when lowering $T$ towards $\Tc$.
    (f-j) show the dependency of $\chiCDW$ on $T$ at $\qvec{}$ points along path R-M.
    (a-b) show the peak of $\chiCDW$ is locked at $(\pi, \pi, \pi)$ when lowering $T$ towards $\Tc$ at $\delta = 0, 10$ percent.
    (c-d) show the peak of $\chiCDW$ drifts along the R-M path to $(\pi, \pi, 5\pi/6)$ and $(\pi, \pi, 4\pi/6)$ respectively, when enlarging $\delta$ to $20\%$ and $30\%$.
    (d) also shows the dome at R point splits off towards adjacent $\qvec{}$ points and multiple cusps are induced, indicating stronger charge density fluctuation.
    (e) shows there is no divergence of $\chiCDW$ even down to $T=0.01$.
    (f-g) show the precedence of $(\pi, \pi, \pi)$ pattern remains until $\Tc$.
    (h) shows the fierce competition among charge density correlations of ordering vector $(\pi, \pi, \pi)$, $(\pi, \pi,5\pi/6)$ and $(\pi, \pi, 4\pi/6)$.
    (i) shows the hot spot of competition drifts along the R-M path, happening between charge density correlations of ordering vector $(\pi, \pi,4\pi/6)$ and $(\pi, \pi, 3\pi/6)$.
    (j) shows $\chiCDW$ is nearly independent of $T$ and no strong charge density correlation grows up.}
    \label{fig:chi_doping_ph1}
\end{figure*}

We then further examine in detail how CDW emerges under low temperature and how doping induces the transition from c-CDW to i-CDW or even destroys CDW completely.
Fig.~\ref{fig:chi_doping_ph1}(a-e) show the distribution of $\chiCDW$ on path $\Gamma$-X-M-$\Gamma$-R-M above $\Tc$ at $\Omega=1$ with $\delta = 0, 10, 20, 30, 40$ percent.
To clarify the competition among $\qvec$ along the R-M path in reciprocal space, Fig.~\ref{fig:chi_doping_ph1}(f-j) show the dependency of $\chiCDW$ on $T$ at different $\qvec{}$ on path R-M under the corresponding $\delta$.
At half-filling [Fig.~\ref{fig:chi_doping_ph1}(a)], $\chiCDW$ is nearly independent of $\qvec{}$ and suppressed to near zero value when $T=2$, where thermal fluctuation destroy any possible charge density correlation.
When lowering temperature to $T=1$, charge density correlation becomes stronger and $\chiCDW$ is enlarged gradually in the entire reciprocal space.
But only at M and R points do $\chiCDW$ gets domed and the dome at R point is higher.
That means there is a competition between two types of CDW with ordering vectors $(\pi,\pi,0)$ and $(\pi,\pi,\pi)$, and the latter takes precedence.
When finally approaching $\Tc=0.39$, the dome at R point gets sharply peaked while the dome at M point is still only a dome.
So the CDW with $\qmax=(\pi, \pi, \pi)$ wins the competition under $\Tc=0.39$.
The race of different $\qvec{}$ points when lowering temperature is also illustrated in Fig.~\ref{fig:chi_doping_ph1}(f).
At mild filling ($\delta=10\%$) as shown in Fig.~\ref{fig:chi_doping_ph1}(b,g), $\chiCDW$ behaves the same as the half-filling case.
The CDW with ordering vector $(\pi, \pi, \pi)$ is robust against $\delta=10\%$ although $\Tc$ is suppressed to $0.35$.

When doping is increased to $20\%$ [Fig.~\ref{fig:chi_doping_ph1}(c,h)], $\chiCDW$ exhibits domes at M and R points like the half-filling case under $T=0.5$.
The dome at R point shifts $\pi/6$ along R-M path to $(\pi,\pi,5\pi/6)$ when lowering temperature to $T=0.2$.
Only then did we recognize the CDW with $\qmax=(\pi,\pi,5\pi/6)$ as a strong player in the competition among different types of CDW.
When finally approaching $\Tc=0.16$, the dome at $(\pi,\pi,5\pi/6)$ gets sharply peaked with a shoulder at R point.
So the CDW with $\qmax=(\pi,\pi,5\pi/6)$ wins the competition under $\Tc=0.16$.
Figure~\ref{fig:chi_doping_ph1}(h) shows there are three patterns of charge density correlation with ordering vectors $(\pi, \pi, \pi)$, $(\pi, \pi, 5\pi/6)$ and $(\pi, \pi, 4\pi/6)$ strongly compete with each other at $\delta=20\%$.
When lowering the temperature, the charge density correlation with $\qmax=(\pi,\pi,\pi)$ loses the race, and a crossover happens towards the precedence of the $(\pi, \pi, 5\pi/6)$ pattern.

As the doping reaches 30\% [Fig.~\ref{fig:chi_doping_ph1}(d)], there are more charge density correlations with different $\qmax$ joining the competition.
Even at relatively high temperature $T=0.17$, the dome at R point splits into two adjacent shoulders at $(\pi, \pi, 4\pi/6)$ and $(5\pi/6, 5\pi/6, 5\pi/6)$.
When lowering the temperature to 0.07, the dome at $(\pi, \pi, 4\pi/6)$ becomes sharper, while the dome at $(5\pi/6, 5\pi/6, 5\pi/6)$ drifts towards $(4\pi/6, 4\pi/6, 4\pi/6)$.
At $\Tc=0.045$, the dome at $(\pi, \pi, 4\pi/6)$ is sharply peaked and wins the competition.
Meanwhile, some other notable cusps get enhanced at $(4\pi/6, 4\pi/6, 4\pi/6)$, $\Gamma$ and M points, indicating stronger charge density fluctuation.
Figure~\ref{fig:chi_doping_ph1}(i) plots the hot spots along the R-M path, from which one can visualize
the strong competition between $(\pi, \pi, 4\pi/6)$ and $(\pi, \pi, 3\pi/6)$ patterns when lowering the temperature.
Meanwhile, a notable enhance of $\chiCDW$ happens at $(\pi, \pi, 0)$, which is another signal of strong charge density fluctuation.

At $\delta=40\%$ [Fig.~\ref{fig:chi_doping_ph1}(e)], so many types of charge density correlation are induced that the competition is fierce and no one can take precedence.
As a result, all types of charge density correlation suddenly quiet down and there is no tendency of divergence of $\chiCDW$, and hence no CDW transition happens when lowering the temperature.
Only some cusps at $(\pi, \pi, 2\pi/6)$, $(3\pi/6, 3\pi/6, 3\pi/6)$ and $(\pi, 5\pi/6, 0)$ remain.
Figure~\ref{fig:chi_doping_ph1}(j) also shows $\chiCDW$ is suppressed to the order of magnitude of $1$ and nearly independent of $T$ on the R-M path.

\begin{figure}[t]
    \centering
    \includegraphics[width=\linewidth]{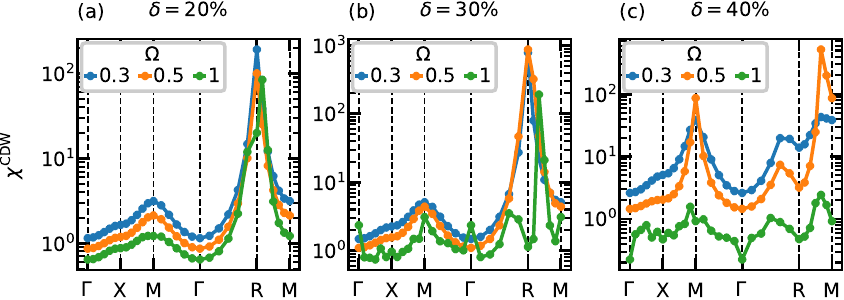}
    \caption{The distribution of $\chiCDW$ on path $\Gamma$-X-M-$\Gamma$-R-M under the temperature just above $\Tc$ at $\Omega = 0.3, 0.5, 1$ with $\delta = 20, 30, 40$ percent.
    Large $\Omega$ can drag $\qmax$ away from $(\pi, \pi, \pi)$ to $(\pi, \pi, 5\pi/6)$ at $\delta=20\%$ [(a)], or even induce more types of charge density correlation at $\delta=30\%$ [(b)] and suppress any CDW transition via flattening $\chiCDW$ at $\delta=40\%$ [(c)].}
    \label{fig:chi_ph_multidoping}
\end{figure}

Figure~\ref{fig:chi_ph_multidoping} shows the distribution of $\chiCDW$ under the temperature just above $\Tc$ at $\Omega=0.3, 0.5, 1$, from which we can further discuss the impact on CDW ordering vector from phonon frequency.
At half filling and $10\%$ doping (SI Fig.~3), the distribution of $\chiCDW$ and hence $\qmax=(\pi, \pi, \pi)$ is robust to $\Omega$.
Upon enlarging doping above $20\%$, phonon manifest its impact on CDW as dragging $\qmax$ away from $(\pi, \pi, \pi)$ and inducing more types of charge density correlation.
At $\delta=20\%$ [Fig.~\ref{fig:chi_ph_multidoping}(a)], lower phonon frequencies lead to the stabilization of the $(\pi, \pi, \pi)$ c-CDW phase while larger phonon frequency $\Omega=1$ drags $\qmax$ to $(\pi, \pi, 5\pi/6)$.
The trend is similar for the $30\%$ doping [Fig.~\ref{fig:chi_ph_multidoping}(b)], but at $\Omega=0.5$ a weak signal of incommensurate charge density correlation 
already manifests itself as a bumped shoulder at $(\pi, \pi, 5\pi/6)$.
When $\delta=40\%$ [Fig.~\ref{fig:chi_ph_multidoping}(c)], the peak of $\chiCDW$ is locked at $\qmax=(\pi, \pi, 2\pi/6)$ and gets sharper when enlarging $\Omega$ to 0.5.
The cusps of $\chiCDW$ at $(4\pi/6, 4\pi/6, 4\pi/6)$ and $M$ point also get sharper.
When further enlarging $\Omega$ to 1, $\chiCDW$ still peaks at $(\pi, \pi, 2\pi/6)$ but the magnitude is significantly suppressed.


\begin{figure*}[t]
    \centering
    \includegraphics[width=\textwidth]{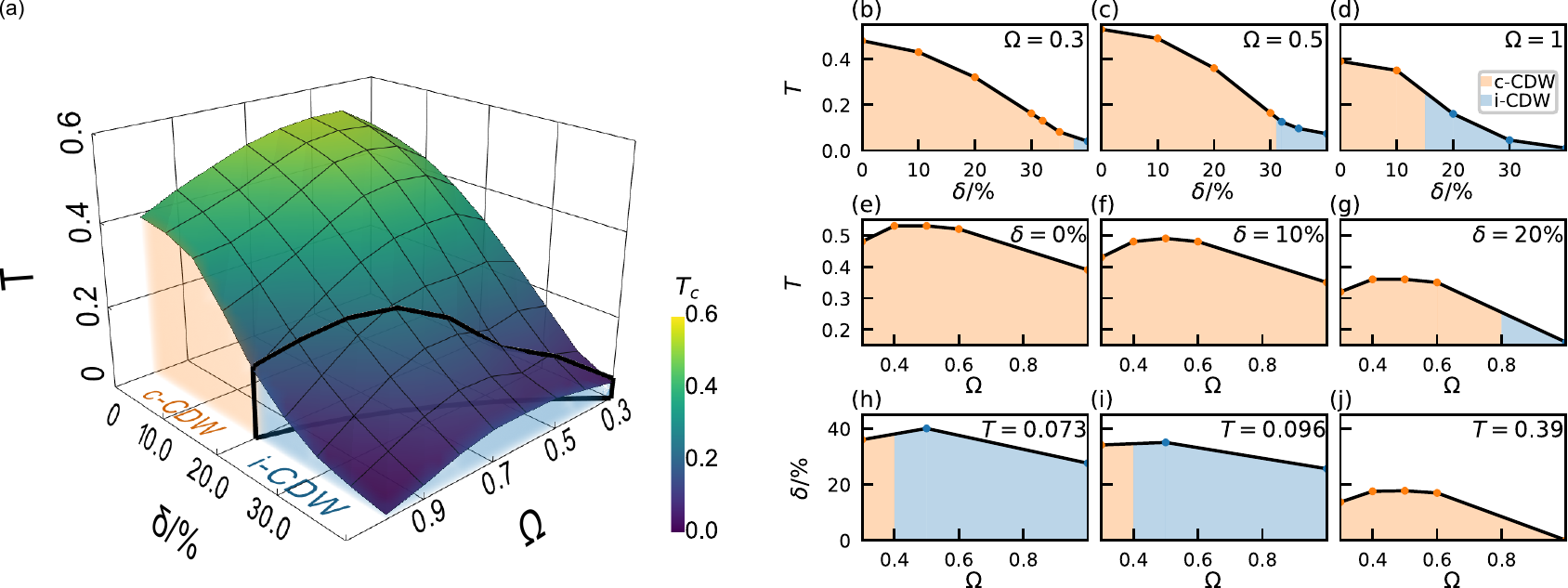}
    \caption{The CDW phase diagram.
    We denote c-CDW and i-CDW as orange and blue respectively.
    (a) is the 3D phase diagram spanned by $\Omega$, $\delta$ and $T$.
    (b-d) are cross sections in $T\text{-}\delta$ plane with $\Omega=0.3, 0.5, 1$, where $\Tc$ gets maximal at half-filling and drops upon hole doping.
    (e-g) are cross sections in $T\text{-}\Omega$ plane with $\delta=0, 10, 20$ percent, where $\Tc$ is parabolic and gets maximal at $\Omega=0.5$.
    (h-j) are cross sections in $\delta\text{-}\Omega$ plane with $T=0.073, 0.096, 0.39$, where $\dcdw$ is parabolic and gets maximal near $\Omega=0.5$.}
    \label{fig:phase_diagram}
\end{figure*}

After determining CDW transition temperature and ordering, we can construct a CDW phase diagram [Fig.~\ref{fig:phase_diagram}(a)] of the cubic Holstein model spanned by $\Omega$, $\delta$ and $T$.
For clarity, we also plot its cross sections in planes of $T\text{-}\delta$ [Fig.~\ref{fig:phase_diagram}(b-d)], $T\text{-}\Omega$ [Fig.~\ref{fig:phase_diagram}(e-g)] and $\delta\text{-}\Omega$ [Fig.~\ref{fig:phase_diagram}(h-j)] respectively.
A common feature at different phonon frequency [Fig.~\ref{fig:phase_diagram}(b-d)] is the monotonic dependency of $\Tc$ on $\delta$.
Near half-filling, CDW is commensurate, and transitions to i-CDW with lower $\Tc$ when enlarging doping.
The higher the phonon frequency, the easier it is for the formation of i-CDW.
On the other hand, the lower the phonon frequency, the more robust c-CDW is against doping.
When we compare different doping [Fig.~\ref{fig:phase_diagram}(e-g)], the phase diagrams show that $\Tc$ is parabolic and always gets maximal at $\Omega=0.5$ regardless of $\delta$.
Fig.~\ref{fig:phase_diagram}(h-j) show the CDW phase diagrams in $\delta\text{-}\Omega$ plane at $T=0.073, 0.096, 0.39$ respectively, derived from linear interpolation of phase boundaries in Fig.~\ref{fig:phase_diagram}(b-d).
The CDW critical hole doping $\dcdw$ is parabolic and always gets maximal near $\Omega=0.5$ regardless of $T$ and the commensurability of CDW.
For low temperature, $T=0.073, 0.096$, CDW is commensurate with $\qmax=(\pi, \pi, \pi)$ at $\Omega=0.3$ and turns into i-CDW with $\qmax=(\pi, \pi, \kappa\pi)$ when enlarging $\Omega$.
For higher temperature, $T=0.39$, CDW is always commensurate with $\qmax=(\pi, \pi, \pi)$ at the considered $\Omega$.

So according to Fig.~\ref{fig:phase_diagram}(e-j), among CDWs at different $\Omega$, the one at $\Omega=0.5$, whether commensurate or incommensurate, is the most robust against heating and hole doping.
Our results also show $\Tc=0.53$ at half-filling and $\Omega=0.4$, close to \(\Tc\approx0.4\) in Ref.~\onlinecite{Cohen2020}.

To conclude, via obtaining the solution of cubic Holstein model in a wide range of parameter space, we establish a thorough understanding of el-ph induced CDW, including the emergence of various CDW phases, the competitions and transitions between different types of CDWs.
Our results are also supported by others' theoretical work.
Kumar and Majumdar \cite{Kumar2005} get similar phase diagram of the same model but with classical phonon, which possesses CDW phase expanded away from half-filling.
In experiment, the insulating phase at half-filling in \(\mathrm{BaBiO}_3\) doped with K \cite{hinks1988synthesis, pei1990structural} is completely suppressed when $\delta \approx 30\%$, while it survives even under heavy doping $\delta \approx 60\%$ when doped with Pb \cite{sleight1975high}, an unexplained puzzle noted by Mattheiss and Hamann \cite{mattheiss1983electronic, mattheiss1988electronic} decades ago.
An important implication of our theoretical results is a possible explanation to the mysterious insulating behavior of barium bismuthates under heavy doping.
Several compelling proposals \cite{weber1987pb, varma1988missing} were made, but none of them is fully satisfactory. 
Our results suggest the \MH{} puzzle can be explained via phonon-induced retarded el-el attraction.
When approaching adiabatic limit by lowering phonon frequency, as does doping with Pb instead of K, el-el attraction becomes more retarded, hence c-CDW phase extends to heavy doping region, explaining the puzzle.
Our work offers a solid foundation to further describe other phenomena of the high temperature superconducting bismuthate in the undoped and doped regions in a unified framework.

\section{Acknowledgments}
We thank C.\ H.\ Pao for helpful discussion.
This work was supported by the National Key R\&D Program of China under
Grant No. 2021YFA1400500, and the Strategic Priority Research Program of the Chinese Academy of Sciences under Grant No. XDB33000000. We are grateful for computational resources provided by the High Performance Computing Platform of Peking University, and the Shanghai Supercomputing Center.

\bibliography{ref}

\end{document}